\documentclass[a4paper] {article}
\usepackage{graphics,color,epsfig}
\def\linadj#1{\normalbaselines
	\multiply\lineskip#1 \divide\lineskip100
 	\multiply\baselineskip#1 \divide\baselineskip100
	\multiply\lineskiplimit#1 \divide\lineskiplimit100 }

\begin{document}

\title{\bf A model density of states for Quarks and Gluons in QGP}

\author{ R. Ramanathan, Agam K. Jha$^+$ and S. S. Singh}

\maketitle
\begin{center}

 Department of Physics, University of Delhi, Delhi - 110007, India.  \\ $^+$Department of  Physics, Kirori Mal College, University of Delhi, Delhi-110007,India.

\end{center}
\linadj{200}

\begin{abstract}
 \large We propose an algebraic form for the density of states of quarks and gluons in a Quark-Gluon Plasma (QGP)fireball in quasi-equilibrium with a hadronic medium as $\rho(k)= \frac {\alpha}{k} + {\beta}k + {\delta}k^{2}$, and determine the parameters $\alpha$, $\beta$ and $\delta$ using Lattice Gauge results on the velocity of sound in QGP. The behaviour of the resulting $\rho(k)$ can be easily compared with the thermodynamic data on QGP that is expected from LHC and other RHIC experiments. Our numerical result shows a linear rise of the value of $\rho(k)$ for $k\sim T \approx 160~ to~180~MeV$,which is significant, and throws light on the evolution of the QGP phase.    \\


PACS number(s): 12.38.Mh; 21.65.+f; 25.75.Ld

\end {abstract}

\vfill
\eject

\maketitle

\large There is high expectation for early availability of data on the behaviour of Quark-Gluon Plasma (QGP) and its thermodynamics from ongoing  Relativistic Heavy Ion Collisions (RHIC) experiments especially at LHC. Though the main theoretical results from Lattice Gauge Computations [1] can be tested and counter checked with data, there is enough opportunity to test semi-phenomenological models of QGP. The thermodynamics of the system can easily be analysed with the parametrisation proposed in the paper.Ever since the suggestion that the QGP can be treated as a gas in thermodynamic quasi-equilibrium with its hadron gas environment, there have been a number of proposals for modelling the gas and the density of states of quarks and gluons in them in the literature [3].In natural units ($\hbar =c =1$), the relativistic momentum '$k$' of the quarks and gluons is the argument of the density of states $\rho(k)$, where the total number of states `$N_{q, g}$' is given by

\begin{equation}\label{3.25}
N_{q,g} = \lambda_{q,g} \int_{0}^{\infty}\rho(k)dk~, 
\end{equation}

where the subscripts q,g refer to quarks and gluons and $\lambda_{q,g}$ refers to the QCD multiplicites for quarks and gluons. we parametrise $\rho(k)$ as 

\begin{equation}
\rho(k)= \frac {\alpha}{k} + {\beta}k + {\delta}k^{2}~,
\end{equation}

where $\alpha$,$\beta$ and $\delta$ are scalar parameters obviously dimensional, to be determined from data on QGP thermodynamics when available, but for the present from Lattice gauge simulation QGP behaviour. This structure (2), is independent of any assumption on the nature of the QGP droplet as in previous literature [3].

As in our previous modeling of the QGP droplet in the hadronic medium [3], we take the hadrons to be majorly made up of pions as the most copious component of the medium.

The Free-energy of the system is made up of the following pieces, namely the free energies $F_{q}$ of the u,d,s quarks, the free energy $F_{g}$ of gluons and the free energy $F_{\pi}$ of the pionic hadrons.

The total free energy $F$ by

\begin{equation}
F = \sum_{i=u,d,s}F_{q_{i}}+F_{g}+F_{\pi} 
\end{equation}

for the system of QGP in quasi-equilibrium with the pionic medium.

As in the literature [3], we have the following expression for the free energy pieces in (3),

 \begin{equation}\label{3.20}
F_{q_{i},g} = \mp T \lambda_{q_{i},g}\int dk \rho(k) \ln (1 \pm e^{-(\sqrt{m_i^2 + k^2}) /T})~,
\end{equation}

 and
\begin{equation}
F_{\pi} = (3T/2\pi^2 )\nu \int_0^{\infty} k^2 dk \ln (1 - e^{-\sqrt{m_{\pi}^2 + k^2} / T})~,
\end{equation}

where we take the multiplicities as usual for quark $\lambda_{q}=6$ and for gluon $\lambda_{g}=8$ and the masses $m_{u}=m_{d}=0$ and $m_{s}=150~MeV$ and $\nu$ is the QGP volume which we take $\sim 1~fm^{3}$ as in previous paper [3]   .

With these ingredients we can compute the various thermodynamic parameters of the system,

    the entropy as usual is $S=-(\frac{\partial F}{\partial T})_{V}$, and the specific heat $C_{V}=-(\frac {\partial^{2}F}{\partial^{2}T})_{V}$.
The velocity of sound in the system is $C_{s}^{2}=\frac {S}{C_{V}}$.

Thus using (4), (5) and (2) in (3) and numerically evaluating the integrals we have an algebraic expression of the form

\begin{equation}
C_{s}^{2}=\frac {\alpha\zeta_{T}^{1} +\beta\zeta_{T}^{2} +\delta\zeta_{T}^{3}}{\alpha\zeta_{T}^{1\prime} +\beta\zeta_{T}^{2\prime} +\delta\zeta_{T}^{3\prime}}
\end{equation}

where $\zeta_{T}$'s are the numerically evaluated integrals in the expressions for $S$ and $C_{V}$.

\begin{figure}
\epsfig{figure=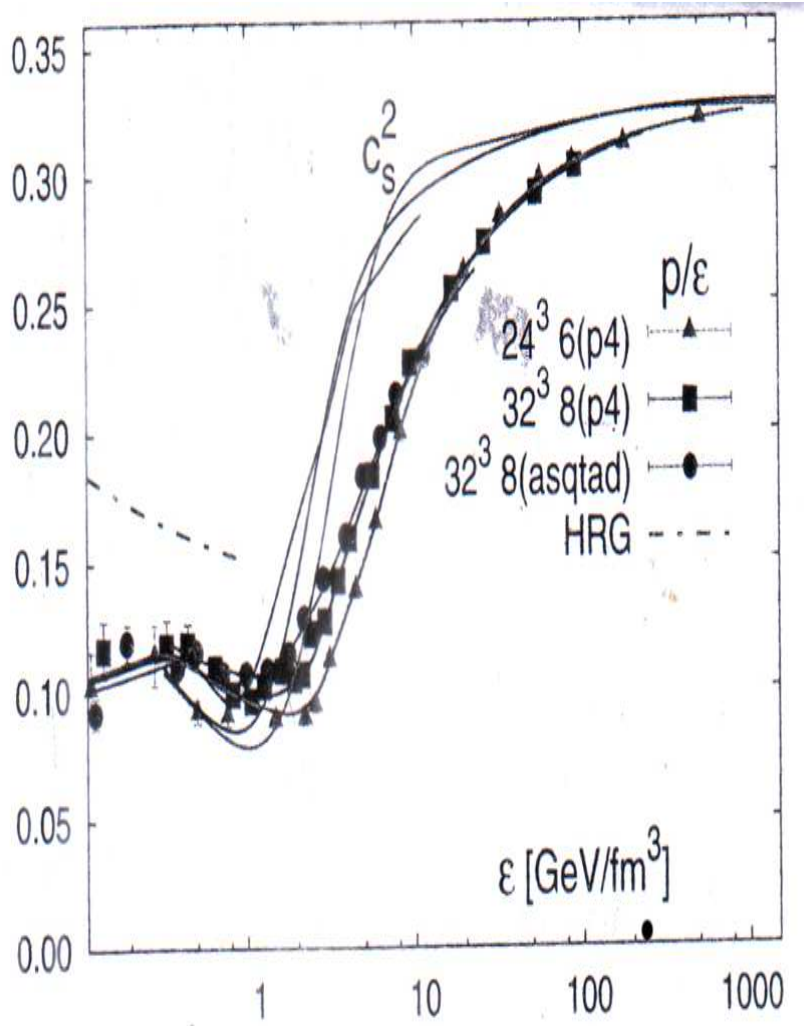,height=5.5in,width=4.5in}
\label{fig15.eps}
\caption{\large  Variation of $C_{s}^{2}$ with energy density [1] .}
\end{figure}

In order to fix $\alpha$,$\beta$ and $\delta$, we choose the random points in the $C_{s}^{2}$ vs. $T$ simulation graph from lattice calculations [1] as shown in Fig.[1] and solve the three simultaneous algebraic equations using the values $f_{T_{1}}$, $f_{T_{2}}$, $f_{T_{3}}$ of $C_{s}^{2}$ at the different temperatures $T_{1}$, $T_{2}$ and $T_{3}$ in relation (6) of the form 

\begin{equation}
\alpha(\zeta_{T_{1}}^{1} - f_{T_{1}}\zeta_{T_{1}}^{1\prime})+ \beta(\zeta_{T_{1}}^{2} - f_{T_{1}}\zeta_{T_{1}}^{2\prime})+ \delta(\zeta_{T_{1}}^{3} - f_{T_{1}}\zeta_{T_{1}}^{3\prime})=0
\end{equation}

and similar equations at temperatures $T_{2}$ and $T_{3}$. We take only the operative region in the graph with $T<300~MeV$ for which RHIC data is expected to be available.

Putting 

\begin{equation}
\zeta_{T_{i}}^{j} - f_{T_{i}}\zeta_{T_{i}}^{j\prime}=\xi_{i}^{j}
\end{equation}

we can recast the three simultaneous equations in (7) as 

\begin{equation}
\left[\begin {array}{ccc}\xi_{1}^{1}&\xi_{1}^{2}&\xi_{1}^{3}\\\noalign{\medskip}\xi_{2}^{1}&\xi_{2}^{2}&\xi_{2}^{3}\\\noalign{\medskip}\xi_{3}^{1}&\xi_{3}^{2}&\xi_{3}^{3}\end {array}\right]\times \left[\begin{array}{c}\alpha \\ \beta \\ \delta \end{array}\right]=\left[\begin{array}{c}0 \\ 0 \\0 \end{array}\right]
\end{equation} 

which can be solved for $\alpha$, $\beta$ and $\delta$ in terms of the known $\xi_{i}^{j}$'s. 

To increase the accuracy, we evaluate $\alpha$, $\beta$ and $\delta$ for a number of random triple points in the $C_{s}^{2}$ vs. T graph with the corresponding $f_{T_{i}}$ values in (9)and then took the average of the values so obtained.

There is also a natural lower bound for $\rho(k)$ in (2) which has a minimum given by a minimum energy value `$k_{m}$' satisfying the equation

\begin{equation}
2\delta k_{m}^{3}+\beta k_{m}^{2}-\alpha=0
\end{equation}

The numerical values are $\alpha=5.7 \times 10^{4}$, $\beta=-0.74~MeV^{-2}$, $\delta=1.3 \times 10^{-2}~MeV^{-3}$ and $k_{m} \equiv 140 ~MeV$.Fig.[2] summaries our result regarding the variation of $\rho(k)$ vs. $k$ dependence and the linear rise of the density of states for $k \sim T$ lies in the band $160~to~180~MeV$ is significant indicating something drastic happening at these temperatures and energies. This feature can be readily checked from the data as soon as it is available, though this only reproduces the result of [1].

\begin{figure}
\epsfig{figure=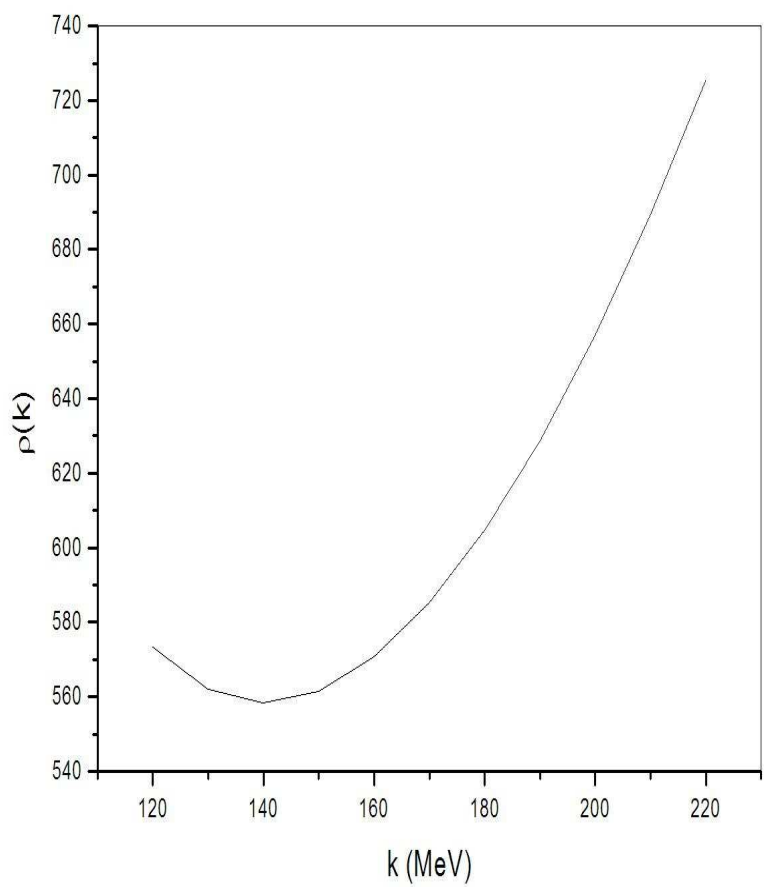,height=5.5in,width=4.5in}
\label{density2.eps}
\caption{\large  Variation of $\rho(k)$ with energy $k$}
\end{figure}

\newpage

{\bf References :}
\begin{enumerate}
\item{P. Petreczky, Nucl. Phys. A 830 (2009).}
\item{F. Karsch, E. Laermann, A. Peikert, Ch. Schmidt and S. Stickan, Nucl. Phys. B (proc. Suppl.) 94, 411 (2001).}
\item{R. Balian and C. Block, Ann. Phys. (N. Y.), 64, 401 (1970); G.Neergaard and J. Madsen, Phys. Rev. D 62, 034005 (2000); R. Ramanathan, Y. K. Mathur,  K. K. Gupta, and Agam K. Jha Phys.Rev.C70,027903 (2004); ; R. Ramanathan, ,  K. K. Gupta, Agam K. Jha, and S.S.Singh, Pramana 68,757 (2007).}

\end{enumerate}

\end{document}